# Delocalized nature of the E'$_\delta$ center in amorphous silicon dioxide

G. Buscarino,*  S. Agnello and  F. M. Gelardi

Department of Physical and Astronomical Sciences, University of Palermo and Istituto Nazionale per la Fisica della Materia, Via Archirafi 36, I-90123 Palermo, Italy

ABSTRACT

We report an experimental study by Electron Paramagnetic Resonance (EPR) of E'$_\delta$ point defect induced by γ ray irradiation in amorphous $SiO_2$. We obtained an estimation of the intensity of the 10 mT doublet characterizing the EPR spectrum of such a defect arising from hyperfine interaction of the unpaired electron with a $^{29}Si$ (I=1/2) nucleus. Moreover, determining the intensity ratio between this hyperfine doublet and the main resonance line of E'$_\delta$ center, we pointed out that the unpaired electron wave function of this center is actually delocalized over four nearly equivalent silicon atoms.

One of the most important reason of interest in amorphous silicon dioxide (a-SiO$_2$) is connected to the fact that a-SiO$_2$ is found as gate in almost the totality of the modern MOS (Metal-Oxide-Semiconductor) devices [1-3]. However, exposing these devices to irradiation, a large concentration of point defects is generated in the oxide layer. Since some of these point defects act as charge traps, the devices undergo a sensible threshold voltage shift that definitively determines their failure [3-4]. The defects prevalently responsible for this effect are the paramagnetic E'$_\gamma$ and E'$_\delta$ centers.

The E'$_\gamma$ center has been widely studied and its most accepted model consists in a positively charged puckered oxygen vacancy: O≡Si$^\bullet$ $^+$Si≡O (where ≡ represents the bonds to three oxygen atoms, $^\bullet$ represents an unpaired electron and $^+$ is a trapped hole) [1, 2, 5, 6], the unpaired electron being localized in a sp$^3$ hybrid orbital of one silicon atom. This structural model followed the definitive attribution to the same defect of a doublet of EPR lines split by ~42 mT, arising from the hyperfine interaction of the unpaired electron with a $^{29}$Si nucleus (4.7% natural abundant isotope with nuclear spin I=1/2) [5]. Following the above reported microscopic model, the E'$_\gamma$ center is considered as the equivalent in a-SiO$_2$ of the E'$_1$ center of quartz [5-11].

The E'$_\delta$ center was observed in X-ray and γ-ray irradiated bulk SiO$_2$ [12-14], in thermally grown thin SiO$_2$ films upon annealing [15-18], and in buried oxide layers obtained by oxygen implantation (SIMOX) [19-20]. The principal EPR characteristics of this center are a main resonance line showing nearly isotropic g tensor (g~2.002) and a pair of line with separation of ~10 mT, supposed to arise from hyperfine interaction of the unpaired electron with a $^{29}$Si nucleus (I=1/2) [12,14]. The hole-trap nature of the E'$_\delta$ center was also verified [20, 21]. An intriguing feature regarding the E'$_\delta$ center is that in the same materials in which this center is induced, another characteristic EPR signal with g~4 is also found, attributed to a point defect in a triplet state (pair of coupled electrons with total spin S=1) [12-14].

The E'$_\delta$ center microscopic structure is still not univocally determined. Until now four distinct models were proposed for this center. The nearly isotropic g tensor, together with the fact that the $^{29}$Si hyperfine splitting (~10 mT) is ~4 times smaller than that of E'$_\gamma$ center (~42 mT), have lead Griscom et al. [12] to introduce a model in which the unpaired electron is delocalized over four mutually orthogonal Si sp$^3$ orbitals, each one similar at all to the one involved in the E'$_\gamma$ center. Since the concentration of E'$_\delta$ center induced by irradiation was found to correlate with the chlorine content of the investigated samples, a model consisting in an electron delocalized over an [SiO$_4$]$^{4+}$ vacancy decorated by three Cl$^-$ ions was proposed (4-Si Cl-containing model). However, as the same authors pointed out, the absence of the EPR lines due to the hyperfine interaction of the unpaired electron with the I=3/2 nuclei of $^{35}$Cl and $^{37}$Cl (with 75.4% and 24.6% natural abundance, respectively) represented a serious difficulty for the reliability of this model. Successively, Tohmon et al. [13] have pointed out that E'$_\delta$ centers can be induced in an equivalent way in Cl- or F-doped SiO$_2$. However, the authors have evidenced that a necessary condition for the formation of this defect is the oxygen deficiency of the material, revealing that the precursor of E'$_\delta$ center is actually an oxygen deficient defect. So, a microscopic model was proposed consisting in an ionized single oxygen vacancy with the unpaired electron nearly equally shared by the two Si atoms (2-Si model). Vanheusden et al. [19] have reported that E'$_\delta$ centers, together with E'$_\gamma$ centers, can be induced in Cl and F free SIMOX materials, clarifying definitively that these impurities are not directly involved in E'$_\delta$ centers. Moreover, an important difference regarding the depth profiles of E'$_\gamma$ and E'$_\delta$ centers was pointed out. The E'$_\delta$ centers, in fact, are induced in the oxide layer nearer to the interface with the Si substrate than the E'$_\gamma$ centers [19]. Basing on this experimental evidence a microscopic structure was proposed for E'$_\delta$ consisting in an unpaired electron delocalized over four Si atoms coordinated to a fifth Si atom disposed at the center of a tetrahedron (5-Si cluster model).

In order to discern between the various models of E'$_\delta$, a relevant role is played by the hyperfine structure. In fact the ratio $\zeta$ between its intensity and that of the main resonance line is expected to have the value:

$$\zeta = \frac{\text{hyperfine doublet EPR intensity}}{\text{main resonance EPR intensity}} = 0.047 \cdot n \cdot (1-0.047)^{(n-1)},$$

where 0.047 is the natural abundance of $^{29}$Si nuclei and n indicate that the unpaired electron wave function is delocalized over n Si atoms. $\zeta$ increases on increasing n because the hyperfine intensity is related to the number of equivalent Si sites of the defect in which the $^{29}$Si nucleus can be found. Zhang et al. [14] reported an experimental estimation of $\zeta$. However, in their samples the concentration of E'$_\delta$ centers was low and consequently the only way to detect the 10 mT doublet was to use the high-power second-harmonic measurements. Nevertheless, by this experimental technique a quantitative estimation of the concentration of centers can not be obtained. So, only *postulating* a strict similarity between the properties of E'$_\gamma$ and E'$_\delta$ centers second-harmonic EPR signals, the authors could estimate $\zeta \cong 0.175$, compatible with the value of 0.163 expected for n=4. This outcome was the basis for the introduction of a microscopic model for the E'$_\delta$ consisting in a [SiO$_4$]$^+$ vacancy (4-Si model).

In this letter we report the first direct experimental estimation of the concentration of defects responsible for the 10 mT hyperfine doublet by ordinary EPR measurements (first-harmonic unsaturated mode). This estimation has permitted us to evaluate the ratio $\zeta$ and to shed new light on the microscopic structure of E'$_\delta$ center.

The material considered in this work is an high purity natural bulk a-SiO$_2$ Type I, *Pursil 453* [22]. An optical absorption band at ~7.6 eV of amplitude larger than 100 cm$^{-1}$, and an absorption band at ~5.0 eV of amplitude ~0.4 cm$^{-1}$ characterize the material as an oxygen deficient silicon dioxide. The Cl and F content of Pursil 453 is lower than ~7x10$^{15}$ cm$^{-3}$ [22].

Samples with size 5x5x1 mm$^3$ were exposed to different γ ray irradiation doses (at room temperature) in the range from 5 kGy to 10$^4$ kGy at a rate ~7 kGy/hr.

EPR measurements were carried out at room temperature and at frequency ν≈9.8 GHz with a Bruker EMX spectrometer working in the first-harmonic unsaturated mode and in high-power second-harmonic mode. In particular, the latter method was used to detect the 10 mT hyperfine doublet when a larger sensitivity was required. All the spectra were acquired with a magnetic-field modulation frequency of 100 kHz. Concentration of defects was determined, with an accuracy of 10%, comparing the double integral of the first-harmonic EPR spectrum with that of a reference sample. For the reference sample the defects concentration was evaluated, with absolute accuracy of 20%, using the instantaneous diffusion method in spin-echo decay measurements carried out in a pulsed spectrometer [23]. The intensity of the second-harmonic EPR signal was estimated by simple integration of the spectra.

In Fig. 1 the EPR spectrum obtained in correspondence to g~2 for a sample irradiated at 10$^3$ kGy is reported. As already pointed out [12], this spectrum arises from the partial superposition of two distinct resonance lines ascribed to E'$_\gamma$ and E'$_\delta$ centers, as indicated by arrows in Fig 1. The EPR signal of the triplet center was also detected in correspondence to g~4 for the same sample. From the analysis of similar spectra obtained for all the other irradiated samples, the dose dependence of the concentration of these three different defects was obtained. E'$_\delta$ and triplet centers were found to grow for doses below ~10$^3$ kGy, at variance for E'$_\gamma$ centers no saturation of concentration was reached up to the highest investigated doses. Second-harmonic measurements were also performed to detect the 10 mT hyperfine doublet but due to the presence of other overlapping signals a quantitative analysis was prevented.

To isolate the 10 mT doublet a sample irradiated at ~$10^3$ kGy was subjected to 25 minutes isochronal thermal treatments in the range of temperature from 330 K to 800 K with step of 10 K. Each temperature of the treatment was stabilized within 3 K. The concentration of E'$_\delta$ was monitored during the thermal treatment and is reported in Fig. 2(a). The E'$_\delta$ center begins to anneal in the temperature range 400 K ÷ 480 K and after that, for T~500 K, a production mechanism is activated and the concentration of defects increases. A maximum concentration ~4 times greater than the initial value is reached after the thermal treatment at ~580 K, while for higher temperature the E'$_\delta$ centers anneal. A growth of concentration was also found for E'$_\gamma$ centers but for these defect the production mechanism was found to start at ~540 K. An analogous growth of concentration is typically observed in irradiated quartz [10], in which for T~500 K the E'$_1$ center concentration grows in correspondence to the annealing of [AlO$_4$]$^0$ hole centers. In the SiO$_2$ sample here considered a concentration of ~$10^{17}$ spins/cm$^3$ of [AlO$_4$]$^0$ centers was estimated by EPR measurements before heating. Furthermore, these impurity centers were found to anneal out in the same temperature range in which the growth of E'$_\delta$ and E'$_\gamma$ centers occur. So, in analogy with the process proposed in quartz [10], we suggest that in Pursil 453 an hole transfer could occur from [AlO$_4$]$^0$ centers to the centers precursors of E'$_\delta$ and E'$_\gamma$.

For temperature of the treatment in the range from 450 K to 650 K the 10 mT doublet can be isolated in the second harmonic spectra, as shown in Fig. 2(b) for T~580 K. The 7.4 mT doublet characteristic of hydrogenated point defect is also distinguishable in this spectrum [24]. Performing a fit procedure with gaussian line shapes for 10 mT and 7.4 mT doublets an estimation of the second-harmonic intensity of the 10 mT pair was obtained. In Fig. 2(a) the intensity of the 10 mT doublet as a function of the temperature of the treatment is shown. From this figure it is evident that the 10 mT doublet temperature dependence is strictly correlated to that of E'$_\delta$ center corroborating the assignment of the doublet to the hyperfine structure of E'$_\delta$ center.

Since the concentration cannot be evaluated by second harmonic measurements alone, first-harmonic unsaturated spectra were also performed for the 10 mT doublet. In Fig. 2(c) (noisy line) the spectrum for the right component of the doublet is reported for T~580 K. We note that partially superimposed to the signal of the 10 mT line, on the low field side of the spectrum, are some structures that vanish for magnetic field higher than ~353.5 mT. Since from second-harmonic measurements we verified that each line of the 10 mT doublet is well described by a gaussian profile, to evaluate the intensity of the 10 mT signal we have superimposed a gaussian derivative line to the experimental spectrum (broken line in Fig. 2c). From this intensity the concentration of centers responsible for 10 mT doublet was determined. This analysis was also repeated after the thermal treatments at T=600 K, T=610 K and T=620 K and the obtained concentration is reported in Fig. 3 as a function of E'$_\delta$ center concentration. These data points show a linear correlation, the slope being the ratio $\zeta$. Performing a best fit procedure the value $\zeta$ =0.16±0.02 was obtained. This intensity ratio is consistent with the value $\zeta$=0.163 expected for n=4, unambiguously indicating that the unpaired electron wave function of the E'$_\delta$ center is actually delocalized over four nearly equivalent silicon atoms.

Our result definitively rule out that E'$_\delta$ center could consist in a ionized single oxygen vacancy. This structural model has been supported by several computational works [25-32] basing on a predicted $^{29}$Si hyperfine doublet compatible with that of the E'$_\delta$ center [25, 26, 30-32]. However, the expected value of $\zeta$ for this defect is 0.089 in disagreement, beyond any experimental uncertainty, with the here reported estimation. So, if this defect really exist in a-SiO$_2$, it should be well distinguishable from E'$_\delta$ center. Furthermore, we note that, owing to its axial symmetry, the EPR signal of the ionized single oxygen vacancy should be different from that of E'$_\delta$ center and more similar to that of E'$_\gamma$.

The models compatible with our data are the 4-Si (Zhang et al. [14]) and the 5-Si cluster (Vanheusden et al. [19]). In both models [14, 19] the E'$_\delta$ center consists in an unpaired electron delocalized in a wave function composed by the four sp$^3$ hybrid orbitals of the nearby Si atoms. The defect could originate from a radiation induced ionization of a pair of nearby oxygen vacancies (O≡Si—Si≡O) [14] or of a 5-Si cluster [19]. Irradiation removes an electron from one of the Si-Si bonds and after a dynamical relaxation the remaining unpaired electron becomes delocalized over four symmetrically disposed silicon atoms. In this scheme and under the hypothesis of similar precursors for E'$_\delta$ and triplet centers [12-14] we suggest that the latter defect could be generated by double ionization of the E'$_\delta$ center precursor.

In conclusion our data support a structure of E'$_\delta$ center in which the unpaired electron is delocalized over four sp$^3$ hybrid orbitals of nearby Si atoms. We stress that this structure agrees with the main experimental evidences of this defect as described in the following. The g tensor is nearly isotropic as expected for delocalized highly symmetric electronic wave function. The hyperfine splitting of the E'$_\delta$ center is ~4 times smaller than that of E'$_\gamma$ center (10 mT≈1/4 · 42 mT) due to delocalization of the electron over four orbitals similar to the one of E'$_\gamma$ center. The intensity ratio ζ between the 10 mT hyperfine doublet and the E'$_\delta$ main EPR line is ζ~0.16. This is the consequence of the existence of four nearly equivalent sites of the defect in which the $^{29}$Si can be localized. Finally, the different depth profiles of E'$_\delta$ and E'$_\gamma$ centers observed in SIMOX samples [19] are a direct consequence of the higher oxygen deficiency needed to the formation of the precursors of E'$_\delta$ (two close oxygen vacancies or small Si cluster) with respect to E'$_\gamma$ centers (single oxygen vacancy). The E'$_\delta$ and E'$_\gamma$ centers are then useful probes to characterize the degree of oxygen deficiency in Si/SiO$_2$ interfaces.

We thank R. Boscaino, M. Cannas, M. Leone, F. Messina for useful discussions and suggestions, E. Calderaro and A. Parlato for taking care of the γ irradiation in the irradiator IGS-3 at the Nuclear Department of Engineering, University of Palermo. This work was financially supported by Ministry of University Research and Technology.

* e-mail address: buscarin@fisica.unipa.it (Gianpiero Buscarino)

FIGURE CAPTION

FIG. 1. First-harmonic unsaturated mode EPR spectrum showing partially superimposed E'$_\gamma$ and E'$_\delta$ resonance lines.

FIG. 2. Pursil 453 sample irradiated at a γ ray irradiation dose of ~$10^3$ kGy: (a) E'$_\delta$ center concentration (left vertical scale) and 10 mT doublet intensity (right vertical scale) as a function of the thermal treatment temperature; (b) high-power second-harmonic EPR spectrum and (c) first-harmonic unsaturated EPR spectrum (noisy line), after the thermal treatment at ~580 K. In (c) superimposed to the spectrum is a derivative gaussian line shape (broken line).

FIG. 3. Concentration of defects responsible for the 10 mT doublet as a function of E'$_\delta$ concentration. The line is obtained by linear fit to the data.

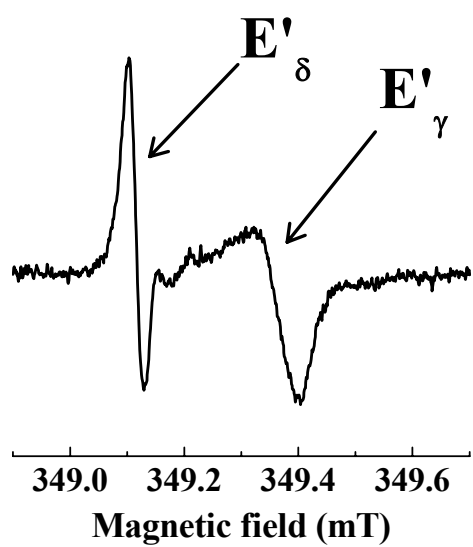

FIGURE 1

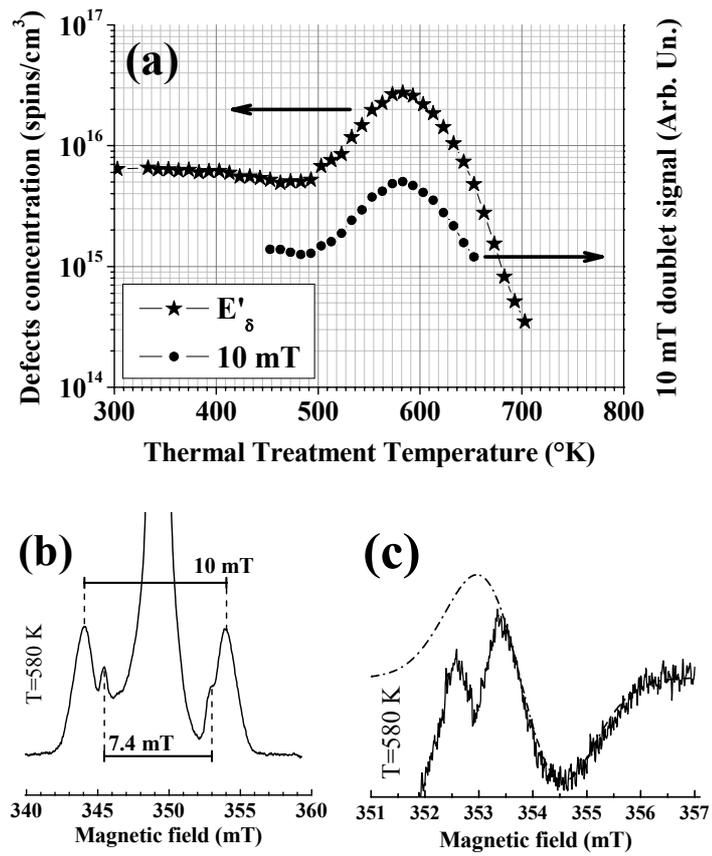

FIGURE 2

G. BUSCARINO et al.

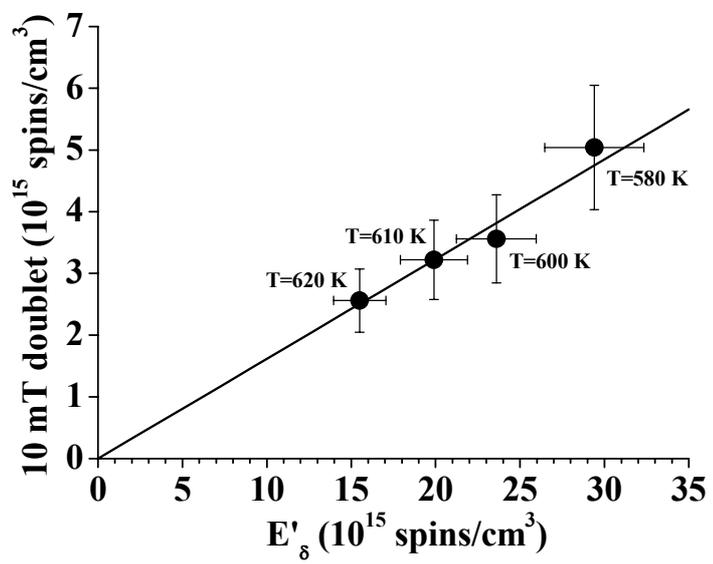

FIGURE 3

G. BUSCARINO et al.